\documentclass[conference]{IEEEtran}

\usepackage{graphicx}
\usepackage{booktabs}
\usepackage[keeplastbox]{flushend}
\usepackage{listings}
\usepackage{color}
\usepackage{lstlinebgrd}
\usepackage{hyperref}
\usepackage{tcolorbox}
\usepackage{lipsum}
\usepackage{tabularx}
\usepackage{tcolorbox,tikz}
\usepackage{changepage}
\usepackage{pgfplots}
\usepackage{pgfplotstable}
\usepackage{filecontents}

\usepackage[inline]{enumitem}

\definecolor{mygreen}{rgb}{0,0.6,0}
\definecolor{mygray}{rgb}{0.5,0.5,0.5}
\definecolor{mymauve}{rgb}{0.58,0,0.82}
\definecolor{diffred}{rgb}{1,0.75,0.75}
\definecolor{diffgreen}{rgb}{0.75,1.0,0.75}

\AtBeginDocument{%
}

\tcbset{colback=white!95!black,colframe=black}

\tcbuselibrary{skins}
\tcbuselibrary{breakable}

\newcommand{\answer}[1]{\begin{tcolorbox}
[breakable,enhanced,colback=white,boxrule=1pt,boxsep=4pt,left=0pt,right=0pt,top=0pt,bottom=0pt,after={\vspace{-0.2cm}}]
{#1}\end{tcolorbox}}

\pgfplotsset{compat=1.13}
\usepgfplotslibrary{statistics}

\begin{document}

\title{Dissection of a Bug Dataset: Anatomy of 395 Patches from Defects4J}

\author{\IEEEauthorblockN{Victor Sobreira\IEEEauthorrefmark{1},
Thomas Durieux\IEEEauthorrefmark{2},
Fernanda Madeiral\IEEEauthorrefmark{1}, 
Martin Monperrus\IEEEauthorrefmark{3}, and
Marcelo de Almeida Maia\IEEEauthorrefmark{1}}
\IEEEauthorblockA{\IEEEauthorrefmark{1}Federal University of Uberl\^andia, Brazil, \{victor, fernanda.madeiral, marcelo.maia\}@ufu.br} 
\IEEEauthorblockA{\IEEEauthorrefmark{2}INRIA \& University of Lille, France, thomas.durieux@inria.fr}
\IEEEauthorblockA{\IEEEauthorrefmark{3}KTH Royal Institute of Technology, Sweden, martin.monperrus@csc.kth.se}
}


\maketitle

\begin{abstract}
Well-designed and publicly available datasets of bugs are an invaluable asset to advance research fields such as fault localization and program repair as they allow directly and fairly comparison between competing techniques and also the replication of experiments. These datasets need to be deeply understood by researchers: the answer for questions like ``which bugs can my technique handle?'' and ``for which bugs is my technique effective?'' depends on the comprehension of properties related to bugs and their patches. However, such properties are usually not included in the datasets, and there is still no widely adopted methodology for characterizing bugs and patches. In this work, we deeply study 395 patches of the Defects4J dataset. Quantitative properties (patch size and spreading) were automatically extracted, whereas qualitative ones (repair actions and patterns) were manually extracted using a thematic analysis-based approach. We found that 1) the median size of Defects4J patches is four lines, and almost 30\% of the patches contain only addition of lines; 2) 92\% of the patches change only one file, and 38\% has no spreading at all; 3) the top-3 most applied repair actions are addition of method calls, conditionals, and assignments, occurring in 77\% of the patches; and 4) nine repair patterns were found for 95\% of the patches, where the most prevalent, appearing in 43\% of the patches, is on conditional blocks. These results are useful for researchers to perform advanced analysis on their techniques' results based on Defects4J. Moreover, our set of properties can be used to characterize and compare different bug datasets.
\end{abstract}



\bstctlcite{MyBSTcontrol}

\section{Introduction}\label{sec:intro}

Bug fixing is a hard and time-consuming task as it involves debugging, i.e., the process of identifying and correcting the root cause of a failure \cite{Zeller2001}. In the last decades, research aiming to automate tasks such as fault localization \cite{Jones2002,Jones2005,Wong2016} and program repair \cite{LeGoues2012genprog,Kim2013par,Mechtaev2016angelix,Martinez2016astor,Xuan2017nopol,Durieux2017,Xiong2017acs} emerged to support developers at fixing bugs. To evaluate the effectiveness of the proposed techniques, researchers either create their own datasets and define ad-hoc baselines or rely on publicly available datasets of bugs (e.g. \cite{Do2005,Dallmeier2007,Just2014,LeGoues2015,Tan2017}. The latter is essential to advance those research fields as publicly available datasets allow directly and fairly comparison between competing techniques and also the replication of experiments.

Researchers in fault localization and program repair fields need detailed information on the datasets they use:
1) to only select bugs that have the required properties according to the technique under consideration (sampling and inclusion criteria), and
2) to perform advanced analysis of the performance of the newly proposed techniques depending on certain properties of the bugs or patches (correlation analysis).

We focus on the analysis of Defects4J \cite{Just2014}, a dataset containing 395 real bugs collected from six open-source Java projects. Although extensively used in recent research on fault localization \cite{BLe2016,Laghari2016,Pearson2017} and program repair \cite{Martinez2016experiment,Le2016HDRepair,Martinez2016astor}, Defects4J does not come with fine-grained information about bugs and their patches.
We contribute to Defects4J with the extraction and study of both quantitative (e.g. metrics) and qualitative properties (e.g. patterns) regarding patches.
This new data is very valuable to
1) interpret past published results based on Defects4J under the light of the extracted properties;
2) provide and guide future research using Defects4J with fine-grained information;
3) understand the representativeness of different kinds of patches in Defects4J to suggest improved versions or new datasets; and
4) compare existing or future datasets of bugs with Defects4J.

We answer to four research questions based on the quantitative and qualitative properties:

\smallskip
\noindent\emph{RQ \#1}: Patch size by number of added, removed and modified lines;

\noindent\emph{RQ \#2}: Patch spreading by number of lines between chunks and by number of modified files, classes and methods;

\noindent\emph{RQ \#3}: Prevalence of repair actions over code elements (e.g. method call addition);

\noindent\emph{RQ \#4}: Prevalence of repair patterns (e.g. wraps-with \texttt{if}).
\smallskip

This study has important implications for future research on program repair, in particular:
there is a need for techniques that leverage patches only containing addition of code;
there is a research avenue for repair algorithms that are specific to repair patterns.


\medskip
To sum up, our contributions are:
\begin{itemize}
\item The \textit{anatomy of the patches in Defects4J} containing an extensive set of patch properties, consolidated into a JSON file\footnote{\url{https://github.com/program-repair/defects4j-dissection}} and augmented with a web user-interface to facilitate exploration\footnote{\url{http://program-repair.org/defects4j-dissection/}};
\item A \textit{bug dataset dissection methodology} to extract valuable quantitative and qualitative properties regarding patches from bug datasets. The methodology is based on diff and advanced patch analysis and combines automated and manual thematic analysis;

\item A \textit{taxonomy of repair actions and patterns}, resulted from manual analysis of patches according our methodology.

\end{itemize}

The remainder of this paper is organized as follows.
\autoref{sec:methodology} presents our methodology, including research questions and data collection.
\autoref{sec:results} presents the answers to the research questions with results and analysis. 
\autoref{sec:lessons} presents lessons learned and
\autoref{sec:threats} discuss threats to validity.
\autoref{sec:rel-works} presents the related work, and
\autoref{sec:conclusion} presents the conclusions.

\section{Methodology}\label{sec:methodology}

In order to characterize and understand patches in Defects4J, we defined the following research questions.

\medskip
\noindent\textbf{RQ \#1: }\textit{What is the size distribution of Defects4J patches?}
Patch size can help to quantify the complexity and difficulty to fix a bug. Small patches have a great potential for repair automation and, in fact, many techniques for automatic program repair works well when this kind of patch is required. The size of human-written patches to fix bugs of a dataset is an important property to know on which bugs a given technique may succeed or not.

\medskip
\noindent\textbf{RQ \#2: }\textit{To what extent are Defects4J patches spread in source code?}
Bug fixing ranges from a single line to multiple lines, and it can be sequential or spread over methods, classes and files. We consider three types of spreading: number of chunks, spreading of chunks and number of modified files, classes and methods. Similar to patch size, patch spreading gives insights on how well a given technique can handle bugs in a dataset used to evaluate such technique. Some past bug datasets, especially those artificially generated, contain many single line bugs (e.g. Siemens suite \cite{Hutchins1994}, available in SIR \cite{Do2005}) and support studies from past to fewer years ago such as \cite{Zhang2014}. There is no guarantee that techniques effective on those bugs are also effective on multi-line spread bugs. Therefore, information about this diversity is essential to sustain any claim based on a specific bug dataset.

\medskip
\noindent\textbf{RQ \#3: }\textit{What is the composition of Defects4J patches in terms of repair actions (i.e. addition, removal and modification) over code elements (e.g. conditional and method call)?}
Actions required on code elements to produce a patch can be simple as a single change of a relational operator, or complex as the addition, removal, and modification of several lines of code, on different code elements, in multiple points in the source code. To proceed with a bug fixing, it is important to know whether a given technique can handle just simple cases as the former or whether such technique is elaborated enough to handle the latter case. Exposing this information from a bug dataset highlights gaps and improvement opportunities, and also avoids misjudgements on a technique being able to handle any type of bug.

\medskip
\noindent\textbf{RQ \#4: }\textit{What repair patterns can be found in Defects4J using a manual thematic analysis \cite{Braun2008}?}
Many patches share some common structures \cite{Martinez2013}.
We are interested in identifying abstractions occurring recurrently in patches that can involve compositions of repair actions. These abstractions shared by groups of patches are called in this paper as \textit{repair patterns}. Knowledge on what repair patterns are present in a dataset can help to develop techniques capable to handle certain types of bugs, for instance by extracting templates for code synthesis.

\subsection{Subject Dataset: Defects4J}

Defects4J \cite{Just2014} is a dataset containing 395 real bugs (version 1.1), built to support software testing research. Bugs in Defects4J were collected from six open-source Java projects: JFreeChart (26 bugs), Closure Compiler (133 bugs), Apache Commons Lang (65 bugs), Apache Commons Math (106 bugs), Mockito Testing Framework (38 bugs) and Joda Time (27 bugs). For each bug, Defects4J delivers the buggy program version and its associated fixed version. Moreover, Defects4J bugs are 1) related to source code (i.e. fixes within the build system, configuration files, documentation, or tests are not included), 2) reproducible (each bug contains at least one test that exposes the bug), and 3) isolated (patches do not include unrelated changes to the bugs such as features or refactorings). In this paper, to refer to Defects4J patches, we use a simple notation with project name followed by bug id, for instance Closure-12 and Math-3.

\subsection{Data Collection}

For each bug, we first produced a \textit{diff} view between the buggy program version and its associated fixed version. We used these views as source for data extraction and analysis, and they are available through hyper-links when patches are cited in this paper. Data collection procedures to answer each research question are described below.

\subsubsection{Patch Size}

We produced scripts to compute how many source code lines were added, removed or modified by a patch, based on the \textit{diff} views.
Addition and removal of lines vary between 
1) consecutive lines 
(Chart-\href{http://program-repair.org/defects4j-dissection/#!/bug/Chart/3}{3}) 
and sparsed lines 
(Chart-\href{http://program-repair.org/defects4j-dissection/#!/bug/Chart/2}{2}), and
2) full statements 
(Closure-\href{http://program-repair.org/defects4j-dissection/#!/bug/Closure/80}{80}), 
partial statements (i.e. closing brackets as in Chart-\href{http://program-repair.org/defects4j-dissection/#!/bug/Chart/26}{26}) and 
line continuation 
(Closure-\href{http://program-repair.org/defects4j-dissection/#!/bug/Closure/59}{59}).
Lines are considered modified when sequences of removed lines are straight followed by added lines (or vice-versa). Thus, to count each modified line, a pair of added and removed lines is needed. 
\autoref{code:Closure-40} shows an example of patch with one modified line (line 635), two non-paired removed lines (the old 636 and 639 lines), and none non-paired added line. By summing these lines, we have the metric \textbf{patch size} in number of lines, which in the example is 3 lines.


\subsubsection{Patch Spreading}

We calculated five metrics of patch spreading also through scripting.
The first metric is \textbf{number of chunks} in a patch. A chunk is a sequence of continuous changes in a file, consisting of the combination of addition, removal, and modification of lines. A patch can be composed of one or more chunks, and this information can give us insights on how a patch is spread through the source code: a patch with a single chunk has no spreading, and the more chunks, the more the patch is spread. 
\autoref{code:Closure-40} has two chunks: the first one is composed of the lines 635 and 636, and the second one is composed of the old line 639.
The second metric is \textbf{spreading of chunks} in a patch. To measure chunk spreading, we consider the number of lines interleaving chunks in a patch. In a patch with only one chunk, this value is naturally zero, because it represents a continuous sequence of changes. In a patch with two chunks, at least one line separates the chunks. For more chunks, naturally, this value tends to increase. An exception is for patches involving more than one file. For this case, we sum the spreading of chunks of all files to get the final spreading of the patch. For example, a patch with two modified files has zero spreading if the patch has just two chunks, one in each file. 
In \autoref{code:Closure-40}, between the old line 636 (end of the first chunk) and the old line 639 (beginning of the second chunk) there is only two lines, which is the value of the metric spreading of chunks in this case. 
It is worth to mention that empty and comment lines were discarded for chunk spreading calculations. These lines have no influence on program behavior, and considering them would make more sense for code readability, for example.
The remaining metrics for patch spreading are \textbf{number of modified files}, \textbf{classes} and \textbf{methods}. We consider only source code files.



\begin{lstlisting} [ caption={Patch for bug Closure-\href{http://program-repair.org/defects4j-dissection/\#!/bug/Closure/40}{40}.}, label={code:Closure-40}]
635     - JsName name = getName(ns.name, false);            
    635 + JsName name = getName(ns.name, true);            
636     - if (name != null) {            
637 636   refNodes.add(new ClassDefiningFunctionNode(
638 637       name, n, parent, parent.getParent()));         
639     - }
\end{lstlisting}

\subsubsection{Repair Actions}

\textit{Diff} views were reviewed manually, aiming to characterize their composition in terms of \textbf{repair actions} over code elements. Repair actions are the basic building blocks for patches. For example, 
\autoref{code:Closure-40} has the actions ``modification'' of ``method call parameter value''  (line 635) and ``removal'' of ``conditional (\texttt{if}) branch'' (old lines 636 and 639).
Repair actions provide fine-grained information beyond simple counting of addition, removal and modification of lines. An initial list was produced based on several potential repair actions. This list was augmented with other actions found during subsequent manual analysis of the patches. For each new repair action found, patches were reviewed to update the annotations on them. Repair actions without occurrences in Defects4J were discarded.

\subsubsection{Repair Patterns}

With knowledge about the content of patches acquired in repair action analysis, it was observed a recurrence of more abstract structures in patches, resembling patterns. For example, 
the modified line 635 of \autoref{code:Closure-40} illustrates the repair pattern ``Constant Change'', while the removed lines 636 and 639 illustrate the repair pattern "Unwraps-from \texttt{if}'' (both discussed in \autoref{sec:repair-pattern-results}).
To confirm the existence of these \textbf{repair patterns}, a process based on Thematic Analysis (TA) was conducted. Originally, TA is ``a method for identifying, analyzing, and reporting patterns (themes) within data'' \cite{Braun2008}. TA is a manual analysis that involves six steps: 1) familiarizing with the data (done with many reading and re-reading of patches to understand its composition); 2) identifying initial codes (in our context, a ``code'' is a single repair action); 3) searching for themes (combinations of repair actions re-appearing over many patches were identified, counted and named as repair patterns); 4) reviewing themes (at first glance some found themes appear to be relevant but after passing all patches it was not sustained, other themes were merged because they were similar, some themes were hierarchically organized and some were discarded); 5) defining and naming themes (although many themes were named at early steps, some of them were reviewed and renamed to better reflect their meaning, and criteria to recognize each instance were defined and exposed in \autoref{sec:repair-pattern-results}); 6) producing the report (this paper and complementary online material reflect this step, compiling the main results of this analysis). 

For both repair actions and patterns, two main activities were performed: 1) the identification of existing repair actions and patterns in the patches, which resulted in a taxonomy of repair actions and repair patterns, and 2) the annotation of the patches in Defects4J using such taxonomy. These two activities were performed manually by the first author of this paper. Then, two other authors validated all the annotated patches by reviewing which repair actions and patterns from the taxonomy a given patch contains, and some adjustments were done.

\subsection{Data Availability}

All the data collected on Defects4J with this methodology is consolidated into a JSON file, which is publicly available in an open-science repository\footnote{\url{https://github.com/program-repair/defects4j-dissection}}.
We also have created a web user-interface to present that data for researchers to easily browse, filter, and understand the patches of Defects4J:

\begin{center}
\url{http://program-repair.org/defects4j-dissection/}
\end{center}

The web user-interface is augmented with runtime information on the Defects4J bugs, which consists of the exceptions threw when running (failing) tests on the buggy program versions, and repair tools that have fixed Defects4J bugs.

\section{Results and Analysis}\label{sec:results}

In this section, we present the results and the answers to our four research questions.

\subsection{Size of the Defects4J Patches (\textbf{RQ \#1})}

Patch size results are presented as follows: we first show the repartition and intersection of patches among three sets, i.e. added, removed and modified lines; then the results for each of these sets; and finally the total size of the patches.

\subsubsection{Repartition and Intersection of Patches}

\begin{figure}[t!]
\centering
\includegraphics[scale=0.5]{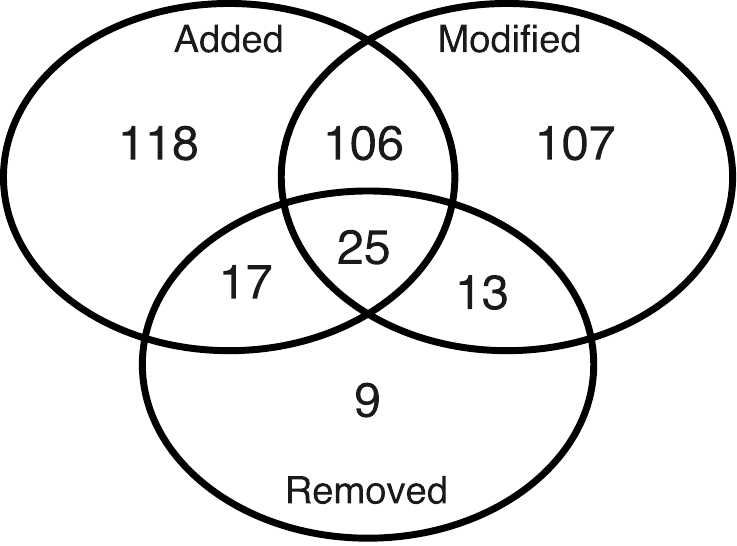}
    \caption{Venn diagram of patches that contain added, removed and/or modified lines.}
\label{fig:LinesAddByProject}
\end{figure}

\autoref{fig:LinesAddByProject} presents the Venn diagram on the intersections among the three types of changes on lines (addition, removal and modification), where each number is the number of patches that contain added, removed and/or modified lines. For example, there are 107 patches containing only modified lines, and there are 106 patches containing both added and modified lines. We note three interesting facts. First, 118 out of 395 patches (29.87\%) contain only added code. This contradicts a common intuition that patches contain mainly code modification. Second, nine patches fix bugs only by removing code, which illustrates that the correct behavior for few cases is already present in the program. 
Third, 234 patches (59.24\%) are exclusive for one of the three types of changes on lines.

\subsubsection{Added Lines}

Addition of lines ranges from 0 to 48 lines as shown in \autoref{tab:LinesDistOverall}.
25\% of the patches have no added line, and at most two lines are added in half of the patches. 
95\% of the patches have added lines ranging from 0 to 19.
Beyond 19 lines, the additions occur in outlier patches.

\subsubsection{Removed Lines}

Numbers on removed lines dropped down considerably when compared with added lines (see the two first lines in \autoref{tab:LinesDistOverall}). For 95\% of the patches, no more than six lines are removed.

\subsubsection{Modified Lines}

As shown in \autoref{tab:LinesDistOverall}, 25\% of the patches have no modified line, half of the patches have at most one modified line, and 95\% of the patches have at most four modified lines.

\subsubsection{Patch Size}

The total size of a patch is the sum of added, removed and modified lines in the patch. As shown in \autoref{tab:LinesDistOverall}, a patch involves at least one line and at most 54 lines.
For 25\% of the patches, at most two lines are involved.
To cover 95\% of the patches, at most 22 lines should be considered.

\answer{
\textbf{RQ \#1: What is the size distribution of Defects4J patches?}\\
\textit{\textbf{Findings:}}
The median size of Defects4J patches is four lines. Addition of lines predominates over removal and modification. In Defects4J, large patches are rare: only 5\% of the patches involve more than 22 lines, and the maximum is 54 lines.\\
\textit{\textbf{Implications:}}
Current repair techniques are only capable of creating small patches. Our results show that Defects4J is an appropriate dataset for program repair because it mostly contains small patches. This confirms the results of \cite{Martinez2016experiment}. Since around 30\% of the patches contain only code addition, research on repair systems should leverage this assumption and define addition-based patches as an important repair search space.
}

\subsection{Spreading of the Defects4J Patches (\textbf{RQ \#2})}

In this section, we analyze the spreading of the patches by the number of 1) chunks, 2) lines between chunks, 3) modified files, 4) modified classes, and 5) modified methods.

\begin{table}[t]
\caption{Descriptive statistics for patch size and spreading.}
\label{tab:LinesDistOverall}
\centering
\setlength{\tabcolsep}{4.5pt}
\begin{tabular}{lrrrrrrrr}
    \toprule
     & Min & 25\% & 50\% & 75\% & 90\% & 95\% & Max \\
    \midrule
    \# Added lines & 0 & 0 & 2 & 6 & 12 & 19 & 48 \\
    \# Removed lines & 0 & 0 & 0 & 0 & 2 & 6 & 24 \\
    \# Modified lines & 0 & 0 & 1 & 2 & 3 & 4 & 27 \\
    Patch size & 1 & 2 & 4 & 9 & 18 & 22 & 54 \\
    \midrule
    \# Chunks & 1 & 1 & 2 & 3 & 5 & 8 & 20 \\
    Spreading & 0 & 0 & 1 & 18.5 & 88.2 & 213.5 & 1,332 \\
	\# Files & 1 & 1 & 1 & 1 & 1 & 2 & 7 \\
	\# Classes & 1 & 1 & 1 & 1 & 1 & 2 & 7 \\
    \# Methods & 0 & 1 & 1 & 2 & 2 & 3 & 20 \\
    \bottomrule
\end{tabular}
\end{table}


\subsubsection{Number of Chunks}

In Defects4J, the patches contain between one and 20 chunks (see \autoref{tab:LinesDistOverall}). Half of the patches contain two or less chunks, and 95\% of the patches contain at most eight chunks.

\subsubsection{Spreading of Chunks}

25\% of the patches in Defects4J have no spreading of chunks.
Half of the patches have a spreading of no more than one line. To cover 95\% of the patches, it is enough to consider spreading of 214 lines (see \autoref{tab:LinesDistOverall}). 
Considering that 95\% of the patches have maximum size of 22 lines and maximum spreading of 214 lines, it is reasonable to state that 95\% of the patches in Defects4J are bounded to blocks of 236 lines overall.

\subsubsection{Modified Files}

92.41\% of the patches modify only one file, and 7.09\% of the patches modify two files. Therefore, for Defects4J bugs, techniques need to access at most 2 files to deal with 99.5\% of the bugs, and if optimized to work with one file, they still cover more than 90\% of the bugs. Just in exceptional cases, patches modify more than two files, which is the case of Mockito-\href{http://program-repair.org/defects4j-dissection/#!/bug/Mockito/19}{19} (five files) and Math-\href{http://program-repair.org/defects4j-dissection/#!/bug/Math/6}{6} (seven files).

\subsubsection{Modified Classes}

We observed the number of modified classes is highly related to the number of modified files in a patch (see \autoref{tab:LinesDistOverall}). Only eight patches modify more classes than files.

\subsubsection{Modified Methods}

We observed two interesting facts when analyzing the number of modified methods in the patches. First, there are two patches that do not modify methods, they only change class and field declaration.
Second, 27\% of the patches (107) change more than one method, and 47\% of these patches are related to the patterns \textit{Copy/Paste} and \textit{Missing Null-Check} (see \autoref{sec:repair-pattern-results}).

\answer{
\textbf{RQ \#2: To what extent are Defects4J patches spread in source code?}\\
\textit{\textbf{Findings:}}
151 patches (38.23\%) are composed of a single continuous chunk. In terms of chunk spreading, 207 patches (52.41\%) have only one code line separating the chunks. Only two patches affect more than two files.\\
\textit{\textbf{Implications:}}
The majority of program repair techniques perform single-point repair by modifying a single location in the code (e.g. GenProg \cite{LeGoues2012genprog}, Nopol \cite{Xuan2017nopol}, Astor \cite{Martinez2016astor}, Elixir \cite{Saha2017}, ssFix \cite{Xin2017}, HDRepair \cite{Le2016HDRepair}, PAR \cite{Kim2013par}). This corresponds well to the 151 single-chunk human-written patches in Defects4J. However, the remaining 244 multi-chunk patches show that there is a need for multi-point program repair, such as Angelix \cite{Mechtaev2016angelix}. As only two patches affect more than two files, single-file fault localization is appropriate for program repair techniques targeting bugs similar to those contained in Defects4J.}

\subsection{Repair Actions in the Defects4J Patches (\textbf{RQ \#3})}

In this section, we present the repair actions over code elements found in the Defects4J patches.

\subsubsection{Assignment}

we consider assignment statements containing the simple assignment operator (\texttt{=}), unary increment (\texttt{x++})/decrement (\texttt{x--}) operators and assignments compound of arithmetic operators (e.g. \texttt{x+=1}). Repair actions related to assignments are:

\begin{itemize}[leftmargin=1.1em]

\item Assignment Addition: 
an assignment to a variable is considered added when it appears in the lines added by the patch and there is no assignment involving such variable in the removed lines. Line 2401 of Closure-\href{http://program-repair.org/defects4j-dissection/#!/bug/Closure/133}{133} shows an assignment addition for the variable \texttt{unreadToken};

\item Assignment Removal: 
an assignment to a variable is considered removed when it appears in the lines removed by the patch and there is no assignment involving such variable in the added lines. Line 145 of Chart-\href{http://program-repair.org/defects4j-dissection/#!/bug/Chart/12}{12} shows the removal of the assignment for the variable \texttt{this.dataset};

\item Assignment Modification: 
an assignment to a variable is considered modified in a patch when its expression changed, i.e., when it appears in the removed and added lines. In Time-\href{http://program-repair.org/defects4j-dissection/#!/bug/Time/7}{7}, the value assigned to the variable \texttt{defaultYear} changed.

\end{itemize}

\subsubsection{Conditional}

constructions considered regarding conditional branches are simple \texttt{if}, \texttt{if-else} (including compact form \texttt{cond?a:b}, i.e. \texttt{if-else-expression}), simple \texttt{else} and \texttt{case} in \texttt{switch} structure. The basic repair actions regarding conditionals are:

\begin{itemize}[leftmargin=1.1em]

\item Conditional Branch Addition: 
Mockito-\href{http://program-repair.org/defects4j-dissection/#!/bug/Mockito/8}{8} has an addition of a simple \texttt{if} in line 79;

\item Conditional Branch Removal: 
Closure-\href{http://program-repair.org/defects4j-dissection/#!/bug/Closure/11}{11} has an \texttt{else-if} branch removal in lines 1314 and 1315.

\end{itemize}

Moreover, the conditional expression can be modified, 
which can happen in three ways:

\begin{itemize}[leftmargin=1.1em]

\item Conditional Expression Modification: Chart-\href{http://program-repair.org/defects4j-dissection/#!/bug/Chart/1}{1} shows an example of a simple change in the conditional expression composition in line 1797;

\item Conditional Expression Expansion: Closure-\href{http://program-repair.org/defects4j-dissection/#!/bug/Closure/99}{99} shows an expansion of a conditional expression composition in line 92;

\item Conditional Expression Reduction: Chart-\href{http://program-repair.org/defects4j-dissection/#!/bug/Chart/5}{5} shows a reduction of a conditional expression composition in line 552.

\end{itemize}

\subsubsection{Loop}

we considerer the loop constructions \texttt{for}, \texttt{while} and \texttt{do-while}. Changes related to these are:

\begin{itemize}[leftmargin=1.1em]

\item Loop Addition: 
addition of a new loop (Closure-\href{http://program-repair.org/defects4j-dissection/#!/bug/Closure/129}{129});

\item Loop Removal: 
removal of an existing loop (Math-\href{http://program-repair.org/defects4j-dissection/#!/bug/Math/56}{56});

\item Loop Modification: 
modification of conditional test (Lang-\href{http://program-repair.org/defects4j-dissection/#!/bug/Lang/19}{19}, line 1050 to 1054) or initialization variables (Math-\href{http://program-repair.org/defects4j-dissection/#!/bug/Math/41}{41}, line 520).

\end{itemize}

\subsubsection{Method Call}

changes related to method calls are present in the majority of the patches. These changes manifest in the following forms:

\begin{itemize}[leftmargin=1.1em]

\item Method Call Addition: 
call addition (Chart-\href{http://program-repair.org/defects4j-dissection/#!/bug/Chart/5}{5}, line 545) 
or parameter addition, i.e., the call is replaced by overloaded version with more parameters (Closure-\href{http://program-repair.org/defects4j-dissection/#!/bug/Closure/3}{3}, line 155);

\item Method Call Removal: 
call removal (Lang-\href{http://program-repair.org/defects4j-dissection/#!/bug/Lang/40}{40}, line 1048) 
or parameter removal, i.e., the call is replaced by overloaded version with less parameters (Math-\href{http://program-repair.org/defects4j-dissection/#!/bug/Math/66}{66}, line 62);

\item Method Call Modification: 
method call replacement (Closure-\href{http://program-repair.org/defects4j-dissection/#!/bug/Closure/4}{4}, lines 190 and 202), method call moving (Closure-\href{http://program-repair.org/defects4j-dissection/#!/bug/Closure/102}{102}, lines 89 and 94), parameter value modification (Lang-\href{http://program-repair.org/defects4j-dissection/#!/bug/Lang/59}{59}, line 884), or parameter value swapping (Time-\href{http://program-repair.org/defects4j-dissection/#!/bug/Time/4}{4}, line 464).

\end{itemize}

\subsubsection{Method Definition}

changes related to method definitions and signatures can be observed by:

\begin{itemize}[leftmargin=1.1em]

\item Method Definition Addition: 
complete method definition addition (Closure-\href{http://program-repair.org/defects4j-dissection/#!/bug/Closure/8}{8}, line 209 to 211) or parameter addition (Closure-\href{http://program-repair.org/defects4j-dissection/#!/bug/Closure/3}{3}, line 280);

\item Method Definition Removal: 
complete method definition removal (Closure-\href{http://program-repair.org/defects4j-dissection/#!/bug/Closure/46}{46}, lines 140 to 155) or parameter removal (Math-\href{http://program-repair.org/defects4j-dissection/#!/bug/Math/66}{66}, lines 94 and 95); 

\item Method Definition Modification: 
method renaming (Mockito-\href{http://program-repair.org/defects4j-dissection/#!/bug/Mockito/21}{21}, line 20),
changes in parameter types (Lang-\href{http://program-repair.org/defects4j-dissection/#!/bug/Lang/30}{30}, 1443 and 1497),
return type (Lang-\href{http://program-repair.org/defects4j-dissection/#!/bug/Lang/29}{29}) and
modifier (Mockito-\href{http://program-repair.org/defects4j-dissection/#!/bug/Mockito/21}{21}, line 20), and changes related to addition and removal of overriding method (Closure-\href{http://program-repair.org/defects4j-dissection/#!/bug/Closure/28}{28}).

\end{itemize}  

\subsubsection{Object Instantiation}

the instantiation of objects is observed by the keyword \texttt{new}, and the changes related to it are:

\begin{itemize}[leftmargin=1.1em]

\item Object Instantiation Addition: 
Mockito-\href{http://program-repair.org/defects4j-dissection/#!/bug/Mockito/36}{36} shows an instantiation addition in line 204;

\item Object Instantiation Removal: 
Math-\href{http://program-repair.org/defects4j-dissection/#!/bug/Math/58}{58} shows an instantiation removal in line 121;

\item Object Instantiation Modification: 
Math-\href{http://program-repair.org/defects4j-dissection/#!/bug/Math/6}{6} shows an instantiation modification in line 51.

\end{itemize}

\subsubsection{Exception}

repair actions related to exception handling and throwing are:

\begin{itemize}[leftmargin=1.1em]

\item Exception Addition: 
addition of \texttt{try-catch} block (Closure-\href{http://program-repair.org/defects4j-dissection/#!/bug/Closure/83}{83}) or \texttt{throw} statement (Time-\href{http://program-repair.org/defects4j-dissection/#!/bug/Time/15}{15}, line 139);

\item Exception Removal: 
removal of \texttt{try-catch} block (Math-\href{http://program-repair.org/defects4j-dissection/#!/bug/Math/60}{60}) or \texttt{throw} statement (Mockito-\href{http://program-repair.org/defects4j-dissection/#!/bug/Mockito/1}{1}).

\end{itemize}

\subsubsection{Return}

repair actions related to \texttt{return} statements are:

\begin{itemize}[leftmargin=1.1em]

\item Return Addition: 
Some patches add a return statement wrapped with an \texttt{if} condition, making them new exit points in the program control flow. Lang-\href{http://program-repair.org/defects4j-dissection/#!/bug/Lang/49}{49} illustrates this case;

\item Return Removal: 
The opposite of the previous case also happens. Closure-\href{http://program-repair.org/defects4j-dissection/#!/bug/Closure/11}{11} shows the removal of a return statement wrapped with an \texttt{if} condition in line 1315;

\item Return Expression Modification: 
Changes in return expression is also common in patches of Defects4J. Math-\href{http://program-repair.org/defects4j-dissection/#!/bug/Math/105}{105} shows an example in line 264.

\end{itemize}

\subsubsection{Variable}

repair actions related to variable declaration and usage are:

\begin{itemize}[leftmargin=1.1em]

\item Variable Addition: 
addition of a new variable declaration (Lang-\href{http://program-repair.org/defects4j-dissection/#!/bug/Lang/40}{40}, lines 1048 and 1049);

\item Variable Removal: 
removal of an existing variable declaration (Math-\href{http://program-repair.org/defects4j-dissection/#!/bug/Math/56}{56}, line 237);

\item Variable Modification: 
modification of the variable type (Chart-\href{http://program-repair.org/defects4j-dissection/#!/bug/Chart/17}{17}, line 857), or modifier (Mockito-\href{http://program-repair.org/defects4j-dissection/#!/bug/Mockito/23}{23}, lines 44 and 45),
or replacement of the usage of a variable by another one (e.g., Lang-\href{http://program-repair.org/defects4j-dissection/#!/bug/Lang/59}{59}, line 884) or by a method call (Math-\href{http://program-repair.org/defects4j-dissection/#!/bug/Math/34}{34}, line 209), preserving the context where it is applied.

\end{itemize}

%

\subsubsection{Type}

\begin{table}[t]
\caption{Repair actions acronyms and grouping names.}
\label{tab:SyntaticStructuresAcronyms}
\centering
\begin{tabular}{ l l l }
  \toprule
  Acronym & Action & Group \\
  \midrule
  asgn & A/R/M & Assignment \\
  cnd  & A/R/M & Conditional \\
  lp   & A/R/M & Loop \\
  mc   & A/R/M & Method Call \\
  md   & A/R/M & Method Definition \\
  obj  & A/R/M & Object Instantiation \\
  ex   & A/R   & Exception \\
  ret  & A/R/M & Return \\
  var  & A/R/M & Variable \\
  ty   & A/M   & Type \\
  \bottomrule
\end{tabular}
\end{table}

patches including changes in types 
are:

\begin{itemize}[leftmargin=1.1em]

\item Type Addition: 
addition of type (Mockito-\href{http://program-repair.org/defects4j-dissection/#!/bug/Mockito/23}{23}, line 136);

\item Type Modification: 
implementation of interface (Math-\href{http://program-repair.org/defects4j-dissection/#!/bug/Math/12}{12}).


\end{itemize}

\pgfplotstableread[col sep=comma, header=true]{data.csv}{\datatable}
\pgfplotstablegetrowsof{\datatable}
\edef\numberofrows{\pgfplotsretval}

\pgfplotsset{%
    discard if not/.style 2 args={
        x filter/.code={
            \edef\tempa{\thisrow{#1}}
            \edef\tempb{#2}
            \ifx\tempa\tempb
            \else
                \def\pgfmathresult{inf}
            \fi
        }
    }
}

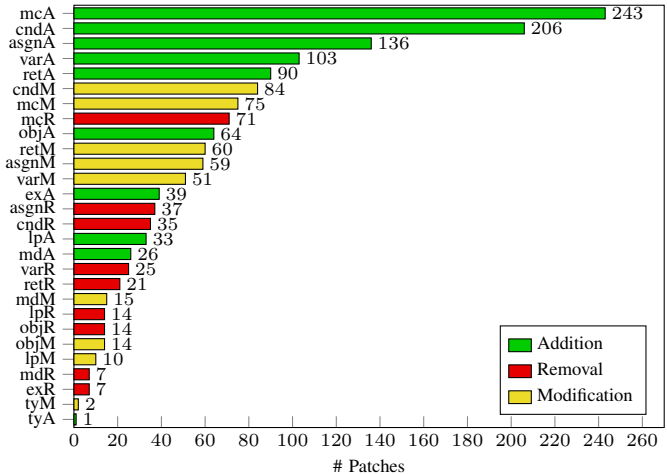
\begin{figure}[t]
  \centering
  \scriptsize
  \begin{adjustwidth}{-1em}{0.5em} 
  \begin{tikzpicture}
    \begin{axis}
    [xbar,
    width=1.05\linewidth,
    height=7.2cm,
    bar width=4.2pt,
    nodes near coords,
    nodes near coords align={horizontal},
    xlabel=\# Patches,
    xlabel near ticks,
    xmin=0,
    xmax=270,
    xtick={0,20,...,260},
    ytick={0,1,...,\numberofrows},
    yticklabels from table = {\datatable}{action},
    yticklabel style={align=right},
    y dir=reverse,
    enlarge y limits=0.02,
    xtick pos=left,
    ytick pos=left,
    legend pos=south east,
    legend cell align={left},
    legend image code/.code={
        \draw [#1] (0cm,-0.1cm) rectangle (0.3cm,0.07cm);
    },
    ]
    \addplot[bar shift=0pt, draw=black, fill=black!20!green] table[discard if not={keyword}{A}, x=frequency, y expr=\coordindex, col sep=comma]{data.csv};
    \addplot[bar shift=0pt, draw=black, fill=black!10!red, discard if not={keyword}{R}] table[x=frequency, y expr=\coordindex, col sep=comma]{data.csv};
    \addplot[bar shift=0pt, draw=black, fill=black!10!yellow, discard if not={keyword}{M}] table[x=frequency, y expr=\coordindex, col sep=comma]{data.csv};
    \legend{Addition,Removal,Modification}
    \end{axis}
  \end{tikzpicture}
  \end{adjustwidth}
  \caption{Incidence of the repair actions in patches.}
  \label{fig:RepairActionsIncidence}
\end{figure}

\autoref{fig:RepairActionsIncidence} shows the ranking of the repair actions over code elements (vertical axis) concerning the number of patches (horizontal axis) where they occur. We grouped repair actions that belong to the same group (e.g. method call and method call parameter addition belong to the group \textit{Method Call Addition}) to avoid a too fragmented graph. We also contracted repair action group names to reduce visual pollution. \autoref{tab:SyntaticStructuresAcronyms} shows, for each group (e.g. \textit{Method Call}), its acronyms (e.g. mc) and suffix letters of the existing action types for it (A=Addition, R=Removal and M=Modification), which combined form the contracted name of a repair action group (e.g. ``mcA'' represents \textit{Method Call Addition}). To make easier to understand the graph, green bars represent addition, red bars represent removal, and yellow bars represent modification actions.

\textit{Method Call Addition} is the most prevalent repair action in the patches (243 patches), followed by \textit{Conditional Branch Addition} (206 patches) and \textit{Assignment Addition} (136 patches). Together and discounting co-occurrences, these three repair actions cover 77.21\% of the patches. In all cases, adding structures surpass removing or modifying existing ones.

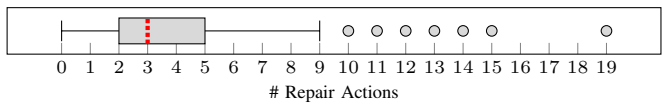
\begin{figure}[b]
  \centering
  \scriptsize
  \begin{tikzpicture}
    \begin{axis}
      [
      width=1.16\linewidth,
      height=2.2cm,
      enlarge y limits=0.4,
      xlabel=\# Repair Actions,
      xlabel near ticks,
      xtick={0,1,...,19},
      xtick pos=left,
      ytick=\empty
      ]
      \addplot[draw=black,fill=black!15,
      boxplot prepared={
        lower whisker=0,
        lower quartile=2,
        median=3,
        upper quartile=5,
        upper whisker=9,
        every median/.style={densely dotted,red,ultra thick},
      }]
      table [row sep=\\,y index=0] { 10\\ 11\\ 12\\ 13\\ 14\\ 15\\ 19\\ };
    \end{axis}
  \end{tikzpicture}
  \caption{Distribution of number of repair actions per patches.}
  \label{fig:Distribution_actions}
\end{figure}

\autoref{fig:Distribution_actions} presents the distribution of number of actions per patches. The median (highlighted in red) shows that 50\% of the patches have no more than three types of repair actions. Some outlier patches were found, containing 10-15 repair actions and a maximum of 19 repair actions.

\answer{
\textbf{RQ \#3: What is the composition of Defects4J patches in terms of repair actions over code elements?}\\
\textit{\textbf{Findings:}}
Addition of method calls, conditionals and assignments are the top-3 applied actions by patches appearing in 305 patches (77.21\%). The additions surpass removals and modifications, which confirms the findings in the RQ \#1 at line level.\\
\textit{\textbf{Implications:}}
Program repair research has mostly focused on conditional statements and assignments \cite{Xuan2017nopol,Nguyen2013semfix,Long2015}. However, handling method calls with rich side-effects has been rather neglected. To our knowledge, only generate-and-validate techniques \`a la GenProg are capable of synthesizing such patches. Assuming Defects4J well represents the distribution of real patches, this calls for more research on repair strategies handling method calls.
}

\subsection{Repair Patterns in the Defects4J Patches (\textbf{RQ \#4})}\label{sec:repair-pattern-results}

\begin{figure}
    \begin{lstlisting}[title={Pattern 1 -- \textit{Conditional Block}.}, label=pattern:1]
// Conditional Block Addition (Lang-45)
616 + if (lower > str.length()) {
617 +    lower = str.length();    
618 + }
// Conditional Block Addition with Return Statement (Closure-5)
176 + if (gramps.isDelProp()) {
177 +    return false;
178 + }
    \end{lstlisting}

    \begin{lstlisting}[title={Pattern 2 -- \textit{Expression Fix}.},label=pattern:2]
// Logic Expression Modification (Chart-1)
1797 - if (dataset != null) {
1797 + if (dataset == null) {
// Logic Expression Expansion (Mockito-34)
106 - if (m instanceof CapturesArguments) {            
106 + if (m instanceof CapturesArguments && i.getArguments().length > k) {
// Arithmetic Expression Modification (Math-80)
1135 - int j = 4 * n - 1;
1135 + int j = 4 * (n - 1);
    \end{lstlisting}

    \begin{lstlisting}[title={Pattern 3 -- \textit{Wraps-with}.},label=pattern:3]
// Wraps-with if (Time-3)
662 + if (years != 0) {
663    setMillis(getChronology().years().add(getMillis(), years));
664 + }
// Wraps-with if-else-exp (Mockito-29)
29 - description.appendText(wanted.toString());
29 + description.appendText(wanted == null ? "null" : wanted.toString());
// Unwraps-from method call (Closure-9)
183 - String moduleName = guessCJSModuleName(normalizeSourceName(script.getSourceFileName()));
184 + String moduleName = guessCJSModuleName(script.getSourceFileName());
    \end{lstlisting}

    \begin{lstlisting}[title={Pattern 4 -- \textit{Single Line}.},label=pattern:4]
// Single Line (Mockito-34)
106 - if (m instanceof CapturesArguments) {            
106 + if (m instanceof CapturesArguments && i.getArguments().length > k) {
    \end{lstlisting}

    \begin{lstlisting}[title={Pattern 5 -- \textit{Wrong Reference}.},label=pattern:5] 
// Wrong Variable Reference (Chart-11)
275 - PathIterator iterator2 = p1.getPathIterator(null);            
275 + PathIterator iterator2 = p2.getPathIterator(null);
// Wrong Method Reference (Closure-10)
1417 - return allResultsMatch(n, MAY_BE_STRING_PREDICATE);
1417 + return anyResultsMatch(n, MAY_BE_STRING_PREDICATE);
    \end{lstlisting}

    \begin{lstlisting}[title={Pattern 6 -- \textit{Missing Null-Check}.},label=pattern:6]
// Missing Null-Check and Non-Null-Check (Chart-15)
1378 +    if (this.dataset == null)  {
1379 +       return 0.0;
1380 +    }
    [...]
2054 +    if (this.dataset != null) {       
2055         state.setTotal(DatasetUtilities.calculatePieDatasetTotal(
2056                  plot.getDataset()));
2057 +    }
    \end{lstlisting}

    \begin{lstlisting}[title={Pattern 7 -- \textit{Copy/Paste}.},label=pattern:7] 
// Copy/Paste of Method Call Replacement (Chart-19)
698 + if (axis == null) {            
699 +    throw new IllegalArgumentException("Null 'axis' argument.");
700 + }
[...]
976 + if (axis == null) {            
977 +    throw new IllegalArgumentException("Null 'axis' argument.");
978 + }
    \end{lstlisting}

    \begin{lstlisting}[title={Pattern 8 -- \textit{Constant Change}.},label=pattern:8]
// Change in string (Closure-65)
1015 - case '\0': sb.append("\\0"); break;
1015 + case '\0': sb.append("\\000"); break;
// Replacement of constant variable (Closure-14) 
767 - cfa.createEdge(fromNode, Branch.UNCOND, finallyNode);            
767 + cfa.createEdge(fromNode, Branch.ON_EX, finallyNode);
    \end{lstlisting}

    \begin{lstlisting}[title={Pattern 9 -- \textit{Code Moving}.},label=pattern:9]
// Code Moving (Closure-13)
126 - traverse(c);            
126   Node next = c.getNext();            
127 + traverse(c);
    \end{lstlisting}

\caption{Code snippet examples for repair patterns.}
\label{fig:patterns}
\end{figure}

We found nine repair patterns from the patches in Defects4J, which are presented in this section in descending order of prevalence. Examples for these patterns are shown in \autoref{fig:patterns}.

\begin{figure*}[ht]
\centering
\vspace{3mm}
\includegraphics[scale=.93,trim=1cm 3.5cm 0.8cm 3.5cm]{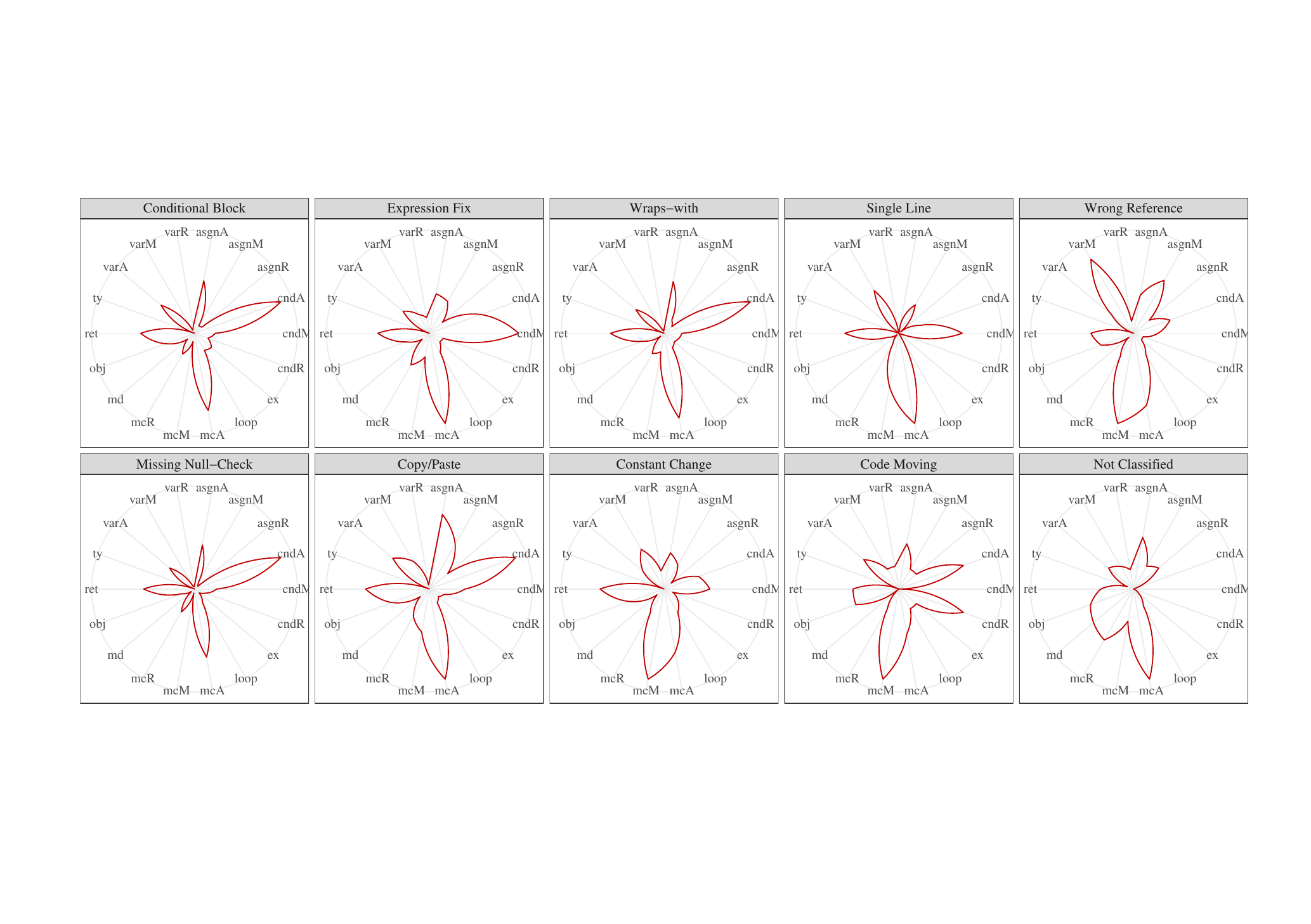}
\caption{Composition of the repair patterns in terms of repair actions.}
\label{fig:PatternsComposition}
\end{figure*}

\subsubsection{Conditional Block}

As shown by the most frequent repair actions in the previous section, the majority of Defects4J bugs require more addition than removing or just changing an existing code. This leads us to the first type of repair pattern, which seems to fill a gap for missing conditional blocks. Following variants of this repair pattern can be found:

\begin{itemize}[leftmargin=1.1em]

\item Conditional Block Addition: involves the addition of a new conditional block (e.g. \texttt{if-then}) in the program (Lang-\href{http://program-repair.org/defects4j-dissection/#!/bug/Lang/45}{45}, lines 616 to 618);

\item Conditional Block Addition with Return Statement: involves the addition of a conditional block that also includes a return statement (Closure-\href{http://program-repair.org/defects4j-dissection/#!/bug/Closure/5}{5}, lines 176 to 178);

\item Conditional Block Addition with Exception Throwing: involves the addition of a conditional block that also throws an exception (Math-\href{http://program-repair.org/defects4j-dissection/#!/bug/Math/48}{48}, lines 189 to 191).


\end{itemize}




There are also patches where conditional blocks are removed as in
Math-\href{http://program-repair.org/defects4j-dissection/#!/bug/Math/50}{50}.



\subsubsection{Expression Fix}

This repair pattern occurs in patches with actions impacting existing logic or arithmetic expressions.

\textit{Logic Expression Fix} mainly occurs in conditional expression (in branches and loops), while \textit{Conditional Block} and \textit{Wraps-with} patterns also impact the code in branch body. Moreover, \textit{Logic Expression Fix} also occurs in return expressions and expressions of assignments to boolean variables, and it has the following variants:

\begin{itemize}[leftmargin=1.1em]
\item Modification: occurs when an existing logic expression is modified, as in Chart-\href{http://program-repair.org/defects4j-dissection/#!/bug/Chart/1}{1}, where the logic operator in the conditional expression was changed;

\item Expansion: occurs when an existing logic expression is preserved and extra logic is added, as in Mockito-\href{http://program-repair.org/defects4j-dissection/#!/bug/Mockito/34}{34};

\item Reduction: occurs when an existing logic expression has a part of the logic removed, as in Closure-\href{http://program-repair.org/defects4j-dissection/#!/bug/Closure/18}{18}.
\end{itemize}

\textit{Arithmetic Expression Fix} mainly occurs in assignment and return statements, as in Math-\href{http://program-repair.org/defects4j-dissection/#!/bug/Math/80}{80} and Math-\href{http://program-repair.org/defects4j-dissection/#!/bug/Math/2}{2}, respectively.

\subsubsection{Wraps-with}
  
This repair pattern resembles the \textit{Conditional Block} pattern, but its intrinsic feature is the wrapping of an existing code with a conditional branch. Indeed, the wrapping structure goes beyond conditionals and can also involve try-catch blocks, method calls and loops. The following are the variants for \textit{Wraps-with}:

\begin{itemize}[leftmargin=1.1em]

\item if: occurs when an existing code is wrapped with a conditional logic using an \texttt{if} expression. Time-\href{http://program-repair.org/defects4j-dissection/#!/bug/Time/3}{3} illustrates this repair pattern applied ten times;

\item if-else: occurs when an existing code is wrapped with an \texttt{if-else} expression, where the existing code can be placed in the \texttt{then} or in the \texttt{else} block. Mockito-\href{http://program-repair.org/defects4j-dissection/#!/bug/Mockito/29}{29} illustrates this repair pattern, where the method call \texttt{wanted.toString()} was wrapped with an \texttt{if-else} expression of type \texttt{cond?a:b};

\item else: occurs when an existing code is wrapped with an \texttt{else} complementing previous written \texttt{if} or \texttt{if-else} (just Chart-\href{http://program-repair.org/defects4j-dissection/#!/bug/Chart/21}{21});

\item try-catch: occurs when an existing code is wrapped with a new \texttt{try-catch} block (Closure-\href{http://program-repair.org/defects4j-dissection/#!/bug/Closure/83}{83});

\item method: occurs when an expression is wrapped with a method call (Math-\href{http://program-repair.org/defects4j-dissection/#!/bug/Math/105}{105});

\item loop: occurs when a statement or block of code is wrapped with a loop, turning it from a simple sequence of code into a repeating one (Closure-\href{http://program-repair.org/defects4j-dissection/#!/bug/Closure/124}{124}).

\end{itemize}

There are cases where \textit{Wraps-with} is more subtle, as in Mockito-\href{http://program-repair.org/defects4j-dissection/#!/bug/Mockito/29}{29}, line 29. It shows the wrapping of the method call \texttt{wanted.toString()} with an \texttt{if-else} expression of type \texttt{cond?a:b}.

The inverse of \textit{Wraps-with} pattern is also found, called \textit{Unwraps-from}. Closure-\href{http://program-repair.org/defects4j-dissection/#!/bug/Closure/9}{9} shows the unwrap of the method call \texttt{script.getSourceFileName()}. Chart-\href{http://program-repair.org/defects4j-dissection/#!/bug/Chart/18}{18} shows the most common case, \textit{Unwraps-from if-else}.

\subsubsection{Single Line}

Patches with the \textit{Single Line} pattern are patches with one line addition, one line removal or both (line modified). We also consider as \textit{Single Line} the special cases when a single statement spans multiple lines (Closure-\href{http://program-repair.org/defects4j-dissection/#!/bug/Closure/55}{55}) or is moving (Closure-\href{http://program-repair.org/defects4j-dissection/#!/bug/Closure/13}{13}). Many of current techniques in fault localization and program repair fields works well where this kind of fixing is required since many of previous datasets are based on seeded faults or mutants \cite{Pearson2017}. These patches involve small repair actions, such as variable replacement by another variable (Chart-\href{http://program-repair.org/defects4j-dissection/#!/bug/Chart/11}{11}) and conditional expression expansion (Mockito-\href{http://program-repair.org/defects4j-dissection/#!/bug/Mockito/34}{34}).


\subsubsection{Wrong Reference}

A wrong reference occurs when a variable or a method call is referenced by mistake instead of another variable or method call. In the patch, the wrong reference is replaced by another one. Examples are:

\begin{itemize}[leftmargin=1.1em]

\item Wrong Variable Reference: Chart-\href{http://program-repair.org/defects4j-dissection/#!/bug/Chart/11}{11} shows the patch for a wrong variable reference, which is replaced by another variable.
In some cases, the wrong variable reference is replaced by a method call such as
Lang-\href{http://program-repair.org/defects4j-dissection/#!/bug/Lang/57}{57};

\item Wrong Method Reference: Closure-\href{http://program-repair.org/defects4j-dissection/#!/bug/Closure/10}{10} shows a wrong method reference, which is replaced by another method reference. In some cases, the wrong method reference is less perceptive (e.g. overloaded method call replacements) such as in Math-\href{http://program-repair.org/defects4j-dissection/#!/bug/Math/70}{70}. In a few cases, the wrong method reference is replaced by a variable (Math-\href{http://program-repair.org/defects4j-dissection/#!/bug/Math/67}{67}).

\end{itemize}

\subsubsection{Missing Null-Check}

This repair pattern is related to the addition of conditional expressions or expansion of existing ones with null-checks. There are examples of \textit{positive null-checks}, where a reference is checked for nullity (line 1378 of Chart-\href{http://program-repair.org/defects4j-dissection/#!/bug/Chart/15}{15}), and \textit{negative null-checks}, where a reference is checked for non-nullity (line 2054 in Chart-\href{http://program-repair.org/defects4j-dissection/#!/bug/Chart/15}{15}).



\subsubsection{Copy/Paste}

Some patches repeat the same change in different points, resembling a copy-paste operation. Chart-\href{http://program-repair.org/defects4j-dissection/#!/bug/Chart/19}{19} shows the addition of the same \textit{Conditional Block with Exception Throwing} in two different methods. Math-\href{http://program-repair.org/defects4j-dissection/#!/bug/Math/71}{71} illustrates a case with non-exact fixing code applied to two different files.

\subsubsection{Constant Change}

This pattern is dedicated to changes of constant values in the code. A constant value is either a literal, i.e. a value fixed in the code, or a constant variable, i.e. a variable with final value that cannot be modified by the program during execution. Closure-\href{http://program-repair.org/defects4j-dissection/#!/bug/Closure/14}{14} shows an example of a constant variable being replaced by another constant variable. For literals, we found patches where string (Closure-\href{http://program-repair.org/defects4j-dissection/#!/bug/Closure/65}{65}), boolean (Math-\href{http://program-repair.org/defects4j-dissection/#!/bug/Math/22}{22}), integer (Lang-\href{http://program-repair.org/defects4j-dissection/#!/bug/Lang/19}{19}), and floating-point number (Mockito-\href{http://program-repair.org/defects4j-dissection/#!/bug/Mockito/26}{26}) were modified.

\subsubsection{Code Moving}

Some patches involve moving code lines around, without extra changes to these lines. Although there are not many examples of this repair pattern, it deserves attention because some patches consist basically of this type of change, and it may consist of single line as in Closure-\href{http://program-repair.org/defects4j-dissection/#!/bug/Closure/13}{13}, or multiple lines as in Closure-\href{http://program-repair.org/defects4j-dissection/#!/bug/Closure/117}{117}.

\autoref{fig:PatternsComposition} shows the overall composition of the repair patterns. Each panel corresponds to one of the nine repair patterns. The radial axis corresponds to the repair actions presented in the previous section that co-occur in patches where the repair pattern was found. For instance, in \textit{Expression Fix} panel, it is clear that \textit{Conditional Modification (cndM)} and \textit{Method Call Addition (mcA)} repair actions are the most prevalent in patches containing the \textit{Expression Fix} pattern, while \textit{Type (ty)} actions 
almost do not appear for this repair pattern. Some repair patterns show similarities, e.g. \textit{Conditional Block}, \textit{Wraps-with} and \textit{Missing Null-Check}. In fact, one of the main differences between \textit{Conditional Block} and \textit{Wraps-with} is the presence of non-patch (or wrapped) code between wraps, which does not influence in repair pattern composition. Furthermore, \textit{Missing Null-Check} is present in many conditionals in patches where \textit{Wraps-with} and \textit{Conditional Block} are involved. Other repair patterns have distinctive silhouette, e.g. \textit{Wrong Reference}, pointing out the more recurrent and distinctive actions in these repair patterns. The 22 patches where no repair pattern was found are shown in last panel (\textit{Not Classified}), which points out that unclassified patches are more related to \textit{Method Call Addition} action.

\autoref{fig:Occurrence-of-the-Repair-Patterns} presents the ranking of the repair patterns (vertical axis) concerning the number of patches (horizontal axis) where they occur.
\textit{Conditional Block} is the most prevalent repair pattern found in the patches, followed by \textit{Expression Fix} and \textit{Wraps-with}. On the other hand, \textit{Constant Change} and \textit{Code Moving} are the less prevalent among the repair patterns.

\pgfplotstableread[col sep=comma, header=true]{patterns_graph.csv}{\datatable}

\begin{figure}[b]
  \centering
  \scriptsize
  \begin{tikzpicture}
    \begin{axis}
    [xbar,
    width=0.95\linewidth,
    height=5.2cm,
    bar width=7pt,
    visualization depends on={x \as \originalvalue},
    point meta={x/395*100},
    nodes near coords={\pgfmathprintnumber{\originalvalue}~~\pgfmathprintnumber{\pgfplotspointmeta}\%},
    nodes near coords align={horizontal},
	xlabel=\# Patches,
    xlabel near ticks,
    xmin=0,
    xmax=220,
    xtick={0,20,...,200},
    ytick=data,
    yticklabels from table={\datatable}{pattern},
    yticklabel style={align=right, text width=1.5cm},
    y dir=reverse,
    enlarge y limits=0.09,
    xtick pos=left,
    ytick pos=left
    ]
    \addplot[draw=black, fill=black!15] table[x=frequency, y expr=\coordindex]{\datatable};
    \end{axis}
  \end{tikzpicture}
  \caption{Incidence of the repair patterns in patches.}
  \label{fig:Occurrence-of-the-Repair-Patterns}
\end{figure}
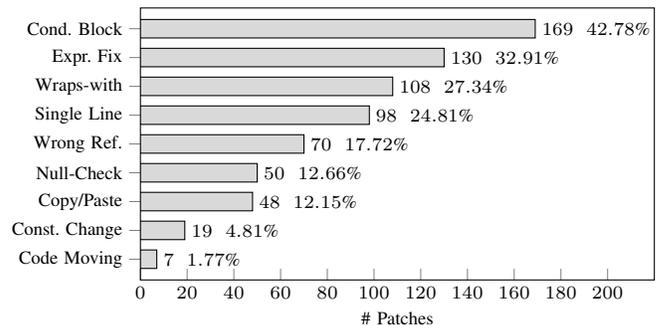

\begin{figure}[h!]
  \centering
  \scriptsize
  \begin{tikzpicture}
    \begin{axis}
      [
      width=1.16\linewidth,
      height=2.2cm,
      enlarge y limits=0.4,
      xlabel=\# Repair Patterns,
      xlabel near ticks,
      xtick pos=left,
      ytick=\empty
      ]
      \addplot[draw=black,fill=black!15,
      boxplot prepared={
        lower whisker=0,
        lower quartile=1,
        median=2,
        upper quartile=2,
        upper whisker=3,
        every median/.style={densely dotted,red,ultra thick},
      }]
      table [row sep=\\,y index=0] { 4\\ 5\\ 6\\ 7\\ };
    \end{axis}
  \end{tikzpicture}
  \caption{Distribution of number of repair patterns per patches.}
  \label{fig:Distribution_patterns}
\end{figure}
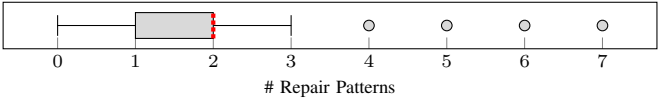

\autoref{fig:Distribution_patterns} presents the distribution of number of patterns per patches. The median (highlighted in red) and the upper quartile are the same, showing that most of the patches (75\%) have no more than two repair patterns. Some outlier patches were found, containing between four and seven repair patterns.

\answer{
\textbf{RQ \#4: What repair patterns can be found in Defects4J using a manual thematic analysis?}\\
\textit{\textbf{Findings:}}
Nine potential repair patterns were identified, which span 373 patches (94.43\%). This show the clear presence of recurring patch techniques. \textit{Conditional Block} repair pattern is the most prevalent, appearing in 169 patches (42.78\%).\\
\textit{\textbf{Implications:}}
Some of the identified repair patterns can guide, for example, the development of program repair tools for specific bug types.
For instance, a variant of the repair pattern \textit{Conditional Block} is the \textit{Conditional (if) Block Addition with Exception Throwing}. It appears in 33 patches and can be synthesized, as confirmed by the recent work on ACS \cite{Xiong2017acs}.
}

\section{Lessons Learned}\label{sec:lessons}

For most bugs in Defects4J, developers added more code than they removed or rewrote existing code. This is first perceived in findings of the RQ \#1, where we studied patch size in terms of added, removed and modified lines. The findings of the RQ \#3 also ground this point when we found that most of repair actions are related to addition of code. 
Researchers and tool builders should also focus on techniques capable to synthesize code, beyond the ones that just modify or remove code, especially for automatic repair. For fault location, the focus should be extended to find the location of missing code and not only to find the location of wrong code (which would mainly lead to modify or remove code).

Delimiting the applicability of a technique by exposing the characteristics of a dataset used to evaluate it is important. For instance, works as \cite{Yokoyama2017} already shown that the performance of a program repair tool changes considerably depending on the types of bug such tool is applied, even for Defects4J projects. Therefore, when reporting results of fault location or automatic repair techniques, it must be clear that not all types of bug would be handled and the dataset used would not contain some bug types, e.g. Defects4J does not cover well bugs spread through many files as pointed out by the findings of the RQ \#2.
Techniques previously evaluated on seeded bugs should be also evaluated on real bugs to confirm their efficacy, as the nature of the former may not reflect the latter. This also shows how fragile can be a technique that relies only on bugs from small or poor designed datasets, not paying attention at overfitting on it. If real bugs are not well represented in the dataset, the technique may be useless in practice. 

The found repair patterns can help to categorize and group patches, leading to a reduced effort while trying to understand what kind of solution can be applied to an existing bug. 
They can help to segment the dataset, avoiding the need to think about a specific solution for every bug.
As pointed out by findings of the RQ \#4, the repair patterns are present in most of the Defects4J patches, and this recurrence may help to optimize the search for fixes with potential to be shared between many bugs.

\section{Threats to Validity} \label{sec:threats}

Although this study provides a deep overview of human patches applied to bugs in six open source projects, our findings may be restricted to the projects in Defects4J. Defects4J can be considered a small dataset of bugs and may not represent well the existing types of bugs and the frequency they occur in the real world. To investigate whether our findings can be generalized to other projects, a follow-up of this work is needed. Although Defects4J projects are Java libraries, they are not necessarily related and, even then, metrics and insights found are consistent between these projects.

The nature of bugs in Defects4J should be also considered when developing solutions for fault location, automatic repair or other applications. Patches in Defects4J are isolated and may not reflect usual commits accompanied by unrelated changes and other kinds of noisy data that an technique should handle. 
Non-source code bugs were not covered in this work, since there is no such type of bug in Defects4J by design. We found two duplicated bugs in Defects4J dataset (i.e., Closure-\{\href{http://program-repair.org/defects4j-dissection/#!/bug/Closure/62}{62} = \href{http://program-repair.org/defects4j-dissection/#!/bug/Closure/63}{63}, \href{http://program-repair.org/defects4j-dissection/#!/bug/Closure/92}{92} = \href{http://program-repair.org/defects4j-dissection/#!/bug/Closure/93}{93}\}). The duplicates were not removed to conduct our analysis and we consider that the impact is negligible over our findings and implications. However, to avoid bias in favor of some bug types, these duplicates may be discarded for other studies or applications.

The taxonomy of repair actions and patterns and the annotation of the patches in Defects4J using such taxonomy are results from manual analysis of patches. Although the patches were carefully analyzed by the first author of this paper and reviewed by other two authors, as any manual work, this one is not free of small mistakes or misinterpretation. Despite the value of manually analyzing human patches, this is a very difficult and time-consuming task. For this reason, by doing this, insights on how to automate the collection of the properties manually extracted from patches were obtained.


We considered all the repair patterns independently in each patch. Since different repair patterns are counted separately, pattern composition in terms of repair actions may be affected by other repair patterns (or other unrelated repair actions) present in the same patch. As a consequence, the identification of a pattern in a patch only implies this pattern is part of such patch. The patch can still involve a much more rich context with other patterns and repair actions beyond an identified pattern. A deeper study on correlations and co-occurrences of patterns would be necessary to obtain an accurate and more detailed description on patterns composition in terms of repair actions (as loosely illustrated in Figure~\ref{fig:PatternsComposition}).
 
\section{Related Works}\label{sec:rel-works}

\subsection{Analysis on Defects4J bugs}

In a recent work, Motwani et al. \cite{Motwani2018} annotated each bug in Defects4J with eleven abstract parameters regarding five defect characteristics: defect importance, complexity, independence, test effectiveness, and characteristics of the human-written patch. One example of abstract parameter is the number of lines edited in a patch, which is used to compute the defect complexity. Similar to our work, they annotated Defects4J bugs with patch size and number of modified files. On the characteristics of the patches, they annotated the bugs with nine code modification types, such as whether the patch contains addition of method calls, which are similar to our repair actions. However, our taxonomy of repair actions is more comprehensive and fine-grained, since we arranged the actions in groups considering more detailed changes. For instance, instead of having the information that a patch changed arguments in a method call, we have the information that an argument was added or removed, or that an argument value was changed or swapped with another one in a method call. Moreover, Motwani et al. considered other information than us, such as the number of relevant test cases, which makes our work and their work complementary to each other.

\subsection{Patch Analysis of Bug Datasets}

Several bug datasets have been proposed to support empirical studies on techniques and tools related to software bugs. Usually, these datasets do not include detailed information on the \textit{bugs} and their \textit{patches} if any (e.g., Siemens suite \cite{Hutchins1994} and SIR \cite{Do2005}), or they include simple information on the \textit{bugs} (e.g., BugBench \cite{Lu2005}), like bug type. In this section, we present notable and recent bug datasets where information about the \textit{patches} are delivered, which is close to our work on Defects4J.

iBugs \cite{Dallmeier2007} (390 Java bugs) contains bugs annotated with size and syntactic properties on their patches. iBugs' size properties include similar patch size and spreading metrics as our work. iBugs' syntactic properties consist of fingerprints describing which syntactic tokens the patch changed, such as keywords, method calls, and expressions, augmented with information on variable usage, operators and literals. These fingerprints are similar to our repair actions, but our taxonomy is organized in a different way. For instance, the groups of token ``keyword'' and ``expression'' in iBugs represent different changes on \texttt{if}; we have the repair action group ``Conditional'' that is dedicated to changes on conditionals. Moreover, our analysis includes repair patterns.

ManyBugs \cite{LeGoues2015} (185 C bugs), besides information on the bugs, delivers manually evaluated information about patches. For each patch, ManyBugs' authors note whenever some changes happened, such as whenever functions, loops, conditional and function calls were added, and whenever arguments to a function or function signature were changed. This is close to our repair actions, but our repair actions are more fine-grained. Similar to our work, they also calculated the number of lines changed (size) and number of files changed (spreading), but different from our work, ManyBugs does not provide number of chunks and repair patterns.

Codeflaws \cite{Tan2017} (3902 C bugs) delivers bugs annotated with syntactic differences between buggy and patch code at AST level. Like in iBugs, Codeflaws' syntactic differences are similar to our repair actions, but we use a more comprehensive taxonomy; for example, in Codeflaws, conditionals and loops are considered together in one group, ``control flow''. Moreover, Codeflaws delivers neither information on patch size and spreading, nor repair patterns.

\subsection{Patch Analysis on Other Resources}

Pan et al. \cite{Pan2009} and Soto et al. \cite{Soto2016} identified patterns in human patches.
Pan et al. \cite{Pan2009} manually analyzed seven open-source projects and found 27 bug fix patterns covering from 46 to 64\% bug fixes. They observed the most common bug fix patterns are related to method call and if condition (both are around 20\% bug fixes), which is consistent with our findings, since \textit{Method Call Addition} and \textit{Conditional Branch Addition} are the most prevalent repair actions in patches.

Soto et al. \cite{Soto2016} focused on identifying how many human patches contain the repair patterns presented by \cite{Kim2013par}. They analyzed 4,590,679 bug fix commits and found that less than 15\% commits contain one of these patterns.
The differences between these two works and this work are the focus on a different dataset, and the collection of additional metrics to characterize human patches, e.g. patch size and spreading.

\section{Conclusions}\label{sec:conclusion}

Research fields related to software bugs require well-designed, publicly available and real bug-based datasets. Defects4J aims at targeting these goals, but lacked an in-depth study to inform the potential users on its contents and characterization of bugs. To fill this gap, we analyzed the anatomy of the Defects4J patches considering four properties: patch size and spreading, and repair actions and patterns. We found that 95\% of the patches in Defects4J involve at most 22 lines, have low spreading in lines (at most 214 lines) and in files (at most two files). Most repair actions involve addition of method calls, conditional branches and assignments (77.21\%). Nine repair patterns were found in 94.43\% of the patches, and \textit{Conditional Block} is the most prevalent one (42.78\%).

Our findings have important implications for those interested in the usage of bug datasets: Defects4J is an appropriate dataset to program repair research; repair based on addition of code should receive as much attention as repair based on modification of code; multi-point program repair is not just a trend but a need for proposed techniques; single file repair responds to the most part of bugs in Defects4J; automatic patch generation can rely on high prevalence of repair actions (e.g. \textit{Method Call Addition}) and patterns (e.g. \textit{Conditional Block}) found. This study is to help researchers to take better and informed decisions around bug dataset choice and comparison.






\section*{Acknowledgement}
We acknowledge CNRS, CAPES (SticAmSud Program), CNPq and FAPEMIG for partially funding this research.






\clearpage

\bibliographystyle{IEEEtran}
\bibliography{IEEEabrv,references}

\end{document}